\documentclass[aps,prl,twocolumn,superscriptaddress,groupedaddress,nofootnoteinbib]{revtex4}  
\usepackage{graphicx}  
\usepackage{dcolumn}   
\usepackage{bm}        
\usepackage{amssymb}   

\def\be{\begin{equation}}
\def\ee{\end{equation}}
\def\ba{\begin{eqnarray}}
\def\ea{\end{eqnarray}}
\def\beq{\begin{eqnarray}}
\def\eeq{\end{eqnarray}}

\def\E{\mathcal{E}}

\def\d{\mathrm{d}}
\def\p{{\cal P}}

\def\K{{\cal K}}

\def\L*{{\cal L}_*}
\def\L{\mathcal{L}}
\def\({\left(}
\def\){\right)}
\def\ie{{\it i.e. }}
\def\nn{\nonumber}
\def\p{\partial}
\def\mn{_{\mu \nu}}
\def\stu{St\"uckelberg }
\def\p{\partial}
\def\mupn{^\mu_{\ \nu}}
\def\<{\langle}
\def\>{\rangle}
\newcommand{\eqref}[1]{(\ref{#1})}

\begin{document}

\title{Ghost free Massive Gravity in the \stu language }

\author{Claudia de Rham}
\address{D\'epartment de Physique  Th\'eorique and Center for Astroparticle Physics, Universit\'e
de  Gen\`eve, 24 Quai E. Ansermet, CH-1211  Gen\`eve}
\address{Department of Physics, Case Western Reserve University, 10900 Euclid Ave, Cleveland, OH 44106, USA}
\author{Gregory Gabadadze}
\address{Center for Cosmology and Particle Physics,
Department of Physics, New York University,
NY, 10003, USA}
\author{Andrew J. Tolley}
\address{Department of Physics, Case Western Reserve University, 10900 Euclid Ave, Cleveland, OH 44106, USA}
\date{\today}


\begin{abstract}

Massive Gravity in four dimensions has been shown to be free of the Boulware-Deser (BD)  ghost in
the ADM language for a  specific choice of mass terms.  We show here how this is consistent with the
\stu language beyond the decoupling limit,  and how the constraint required to remove the BD  ghost arises in
this framework nonperturbatively, without the use of field redefinitions. We emphasize a subtlety
in obtaining  this constraint, that has been overlooked in previous literature. In both the ADM and
\stu formalisms the vanishing of the determinant of a Hessian guarantees the absence of the BD ghost.
\end{abstract}

\maketitle



\section{Introduction}
In recent years there has been considerable progress in the construction of massive gravity theories in 4 dimensions,
\cite{Fierz:1939ix,vDVZ,Arkady,Boulware:1973my,ArkaniHamed:2002sp,Creminelli:2005qk,Gabadadze:2009ja,deRham:2009rm,deRham:2010gu}, culminating in the proposal of \cite{deRham:2010ik,deRham:2010kj} of a two parameter family of models which are ghost free in the decoupling limit and were explicitly shown to be ghost free in the full theory to fourth order, as well as completely nonlinearly in specific cases. Furthermore, it was shown in \cite{Hassan:2011hr}
that the fourth order calculations of \cite{deRham:2010kj} could be generalized to all orders giving a full proof in the ADM formalism
that this class of theories is ghost free.

The central problem faced in any construction of massive gravity is that a naive counting of degrees of freedom would lead to six modes. This can be seen in a number of ways, but in the ADM language it arises because the 3-metric has 6 components.
In the unitary gauge of a massive theory diffeomorphism invariance is lost because of the mass term and so the lapse and shift remain as non-dynamical degrees of freedom which in general do not enforce constraints. As such they can be integrated out, giving a theory of 6 propagating degrees of freedom. Since a massive theory should only have 5, the additional 6th mode is inevitably a ghost and is known as the Boulware-Deser (BD) ghost \cite{Boulware:1973my}.

The resolution to this problem is to construct the mass term in such a way that a constraint remains, which serves to project out nonlinearly the BD ghost. In general the mass term in the ADM form can be expressed as
\ba
S_{\rm mass} = \int \d^4 x \, {\mathcal L}_m(N,N^i,h_{ij})\,.
\ea
The existence of a constraint can be characterized by the statement of the vanishing of the determinant of the Hessian $\det[{\cal H}_{ab}]=0$ where the $4 \times 4$ Hessian ${\cal H}_{ab}$ is constructed as
\ba
{\cal H}_{ab}= \frac{\partial}{\partial N^a}\frac{\partial}{\partial N^b} \left( {\mathcal L}_m(N,N^i,h_{ij}) \right)\,,
\ea
with the shorthand $N^0 = N$.
Whenever the determinant vanishes it implies that it is not possible to solve the equations of motion to determine the lapse and shift directly and so at least one constraint must remain. In practice we require only one additional constraint from this process and so an equivalent way of stating this is that once we have solved the algebraic equations for the shift and substituted back into the action, the resulting action becomes linear in $N$ and so $N$ enforces a new version of the Hamiltonian constraint. When this calculation is done with the phase space form of the action, it is necessary to generate two constraints in phase space to remove one physical degree of freedom. The second constraint simply comes from the commutator of the first constraint with the full Hamiltonian, as it does in the linearized theory.

These arguments describe how in principle the ghost can be removed in a massive theory of gravity in the ADM language. Fortunately it has recently been shown in \cite{Hassan:2011hr} that  the two parameter family described in \cite{deRham:2010kj} satisfies all the above requirements, namely that once the shift is integrated out, the resulting Hamiltonian is linear in the lapse. This is consistent with the argument of \cite{deRham:2010kj} to fourth order, and the all orders argument of \cite{deRham:2010kj} in a two dimensional model.

\section{\stu fields}
The \stu trick has been successful in determining the form of the mass term that will inevitably be ghost free in the decoupling limit. However one can see that away from the decoupling limit, a problem appears to arise in perturbation theory seemingly hinting towards the presence of a ghost. The result would be in clear contradiction with the ADM analysis, and so it behoves us to ascertain where the discrepancy comes from. The essence of the \stu trick is to write the mass term in a gauge invariant manner by defining the covariant tensor
\ba
H_{\mu\nu} = g_{\mu\nu} - \eta_{ab} \partial_{\mu} \phi^a  \partial_{\nu} \phi^b .
\ea
In unitary gauge $\phi^a=x^a$ and so $H_{\mu\nu}=g_{\mu\nu}-\eta_{\mu\nu}$ is just the metric perturbation.
For this to transform as a tensor it is crucial that the $\phi^a$ transform as scalars under diffeomorphisms even though they are vectors under global Lorentz transformations. This point makes it clear that the global Lorentz symmetry is quite separate from the local diffeomorphism invariance. Since this transforms as a tensor we can construct the mass term out of scalar contractions of $H^{\mu}_{\nu}= g^{\mu\alpha}H_{\alpha \nu}$. The key realization of \cite{deRham:2010kj} was that the mass term is inevitably a finite order polynomial in the tensor $\mathcal{K}_{\mu\nu}$ defined such that
\ba
2\mathcal{K}^{\mu}_{\nu}- \mathcal{K}^{\mu}_{\alpha}\mathcal{K}^{\alpha}_{\nu} = H^{\mu}_{\nu}.
\ea
The allowed mass terms derived in \cite{deRham:2010kj}  can be shown to arise as the expansion of the determinant $\sqrt{-g}\det[g^{\mu}_{\nu}+\lambda \mathcal{K}^{\mu}_{\nu}]$ to fourth order in $\lambda$ \cite{Nieuwenhuizen,Hassan:2011vm,Koyama}. With these terms it was shown that no ghosts are present in the decoupling limit.

\subsection{\stu trick in the decoupling limit}

With the \stu fields the absence of a ghost in the decoupling limit has traditionally been determined in one of two ways. One is that it must be possible to write the action in such a way that after integration by parts it is independent of $\dot{\phi^0}$. If this is the case then varying with respect to $\phi^0$ will enforce a constraint which removes the BD ghost. Equivalently it has become conventional to introduce a further \stu field $\pi$ by means of the decomposition $\phi^a = x^a + {\hat V}^a + \partial^a \pi$. With this definition it is clear that $\pi$ is a scalar under global Lorentz transformations. Furthermore $\pi$ captures the physical degree of freedom of the helicity zero mode of the massive graviton. An equivalent way of stating the problem is that the equations of motion for $\pi$ remain second order in time. This amounts to to requiring that up to total derivatives the action can be written in such way that no derivatives higher than $\dot\pi$ enter which is equivalent to the previous criterion.

\subsection{(Apparent) Failure of the \stu trick beyond the decoupling limit}

Although the \stu language has been successfully used to show the absence of ghosts in the decoupling limit, this same logic has been applied in several papers \cite{slavapaper1,slavapaper2} to argue that there is a ghost in massive gravity since it can be shown that when we go away from the decoupling limit terms arise which are quadratic and higher in $\dot{\phi^0}$ in the action, which would lead to the equation obtained by varying the action with respect to $\phi^0$ being dynamical and thus removing this constraint.

However it is easy to see that while the \stu field $\phi^a$ can be treated as a vector in the decoupling limit, this treatment no longer holds beyond the decoupling limit. $\phi^a$ is intrinsically a set of four scalars, and only in the decoupling limit where $\pi$ transforms as a scalar under diffeomorphisms is the previous counting appropriate. Beyond the decoupling limit, there is no unique way to identify the scalar $\pi$, and the previous trick cannot be used. Instead one should carefully check whether the system propagates four independent \stu scalar fields as naively anticipated (in which case the BD ghost would survive) or whether there exist a constraint and not all four \stu fields are dynamical.

Fortunately, it is easy to see that the constraint found in the ADM language does indeed survive in the \stu formalism. There does exist an equation which is obtained from the equations of motion in which neither $\ddot{\phi^0}$ nor $\ddot{\phi^i}$ enters. It is simply that this is not the equation obtained by varying the action with respect to $\phi^0$ but is rather a linear combination of the equations obtained by varying with respect to $\phi^a$. In other words if we denote by $\E_a$ the equation obtained by varying the Lagrangian with respect to $\phi^a$
\be
\E_a = \frac{\delta \mathcal{L}}{\delta \phi^a}=0
\ee
we find that although $\E_0$ contains terms with $\ddot{\phi^0}$, the dynamics of the four scalars are not independent. More precisely, upon inspection, the matrix
\ba
\mathcal{A}_{ab}=\frac{\partial \E_a}{\partial \ddot \phi^b}=\frac{\delta^2 \mathcal{L}}{\delta \dot \phi^a \delta \dot \phi^b}
\ea
has zero determinant and is not invertible.
In particular there exists a linear combination
\be
\mathcal{C} = c^a \E_a=0
\ee
for which $\mathcal{C}$ contains no higher than first order time derivatives of $\phi^a$. Stated in different terms, the Lagrange multiplier for the constraint is not $\phi^0$ but is rather determined by the direction in field space set by $\delta \phi^a \propto c^a$. This means that if we perform the helicity decomposition  $\phi^a = x^a + {\hat V}^a + \partial^a \pi$, it would appear as if the equation of motion for $\pi$ were higher derivative. However these higher derivatives would be removed by using the well defined equations of motion for the $V^a$. Another way of stating this is that at the level of the action it is possible to perform field redefinitions to remove all terms with $\ddot{\pi}$. Whilst this improves the understanding of perturbation theory, it is not necessary to perform these field redefinitions. The presence of a constraint can equivalently be shown nonlinearly in the original variables using the equations of motion.

We shall show how this works explicitly in the next section, however it is worth pausing to ask why the constraint does not arise in the normal manner. The reason can be traced to the fact that whilst $\phi^a$ is a vector under global Lorentz transformations, it is a set of four scalars under diffeomorphisms. The normal logic that varying with respect to $\phi^0$ should determine the missing constraint would naively apply if $\phi^a$ were a vector under diffeomorphisms. For instance this would be an appropriate logic for a massive gauge theory. In the decoupling limit this distinction becomes irrelevant, but in the presence of gravity it becomes extremely important. By working in unitary gauge, the ADM phase space analysis bypasses this peculiarity and gives a clear counting of degrees of freedom. Nevertheless as we shall show in the next section the counting, when done properly, is identical in the \stu picture.


\section{Explicit Example}
\subsection{Locally Inertial Frame}
In this section we will focus solely on the mass term of the Lagrangian.
Even though the absence of ghost in the ADM formalism has been proven for the full two parameter family of mass terms \cite{Hassan:2011hr}, we focus in what follows on the specific potential $\mathcal{K}\mupn\, \mathcal{K}^\nu_{\ \mu}-\mathcal{K}^2$, although the same results holds for any mass term in this class of models.
Since no derivatives of the metric appear in this term, it is convenient to make a local transformation so as to be in the Locally Inertial Frame (LIF) and set the metric locally to $g\mn=\eta\mn$ and Christoffels to zero. Including the metric only alleviates the problem to be
described since the metric would otherwise carry some of the derivatives and in turn reduces the number of derivatives on the scalar fields. By working in the LIF, we therefore ensure to keep the largest possible number of derivatives on the \stu fields. We can then evaluate the equations of motion for the scalars in the LIF and these must be well defined.
To start with, for pedagogical simplicity only, let us consider the specific situation where the four \stu scalar fields are set to $\phi^a=\(\phi^0(t,x),\phi^1(t,x),0,0\)$ (or to be more precise, we focus on a 2-dimensional subclass of models). For that specific configuration, the mass term is then (up to total derivatives and an overall normalization)
\ba
\L_m&=&\mathcal{K}^2-\mathcal{K}\mupn\,  \mathcal{K}^\nu_{\ \mu}\\
&=&\sqrt{(\p_\mu \phi^\mu)^2-F\mn^2}
\ea
with $F\mn=\p_\mu \phi_\nu-\p_\nu \phi_\mu$. It is already clear that due to the term $\p_\mu \phi^\mu$, the Lagrangian is not linear in $\dot \phi^0$, which is potentially worrisome. To make this more explicit, one could split the \stu in a similar way as what is done in the decoupling limit, \ie $\phi^{a}=x^{a}+V^a$ and expand the mass in powers of $V^a$ to obtain
\ba
\L_m&=&-\frac18 \mathcal{F}\mn^2\Big(1-\frac 12 \p_\mu V^\mu\\
 &&+\frac{1}{32}\mathcal{F}\mn^2+\frac{1}{4}(\p_\mu V^\mu )^2 +\cdots\Big)\nn \,,
\ea
with now $\mathcal{F}\mn=\p_\mu V_\nu-\p_\nu V_\mu$.
Now it is straightforward to see that {\it if $V^0$ was the $0^{\rm th}$ component of a vector}, the last term $(\p_\mu V^\mu )^2$ would be problematic as it makes $V^0$ dynamical. Saying it differently, if we were to write $V^0=\dot \pi$ then the equations of motion would involve a term of the form $\dddot \pi$, which would naively manifest the presence of a ghost. If this conclusion was true, it would suggest an inconsistency with the ADM analysis \cite{deRham:2010kj,Hassan:2011hr} which showed the absence of ghosts in the theory. However both results agree when pushing the \stu analysis a step further and properly counting the number of degrees of freedom, as we show explicitly below.

As emphasized earlier, $\phi^\mu$ is not a vector but rather a combination of scalars, so
beyond the decoupling limit, it is not straightforward to identify the helicity zero component. In particular just naively performing a decomposition of the form $\phi^a = x^a+ \hat{V}^a + \partial^a \pi$, $\pi$ alone does not capture the helicity zero mode.
Instead, to check the presence or not of the $6^{\rm th}$ BD mode, one should directly count how many fields are physically propagating. To avoid the ghost, only three out of the four \stu fields should be dynamical while the last one should propagate a constraint. However since $\phi^\mu$ is not a vector, there is no reason why $\phi^0$ (or $V^0$) should be the nondynamical degree of freedom. As it happens here, it is actually a combination of all the \stu fields which generates the constraint and reduces the number of degrees of freedom.

To see this explicitly, and in a completely non-perturbative way, let us return to the action for the two \stu fields $\phi^{0,1}$. The field equations of motion are
\ba
\mathcal{E}_\nu&=&\p_\mu F\mupn +\frac 12 \p_\nu \p_\mu \phi^\mu\\
&-&\frac{\p_\mu((\p_\alpha \phi^\alpha)^2-F_{\alpha\beta}^2)}{(\p_\alpha \phi^\alpha)^2-F_{\alpha\beta}^2}
\(\frac 12 \delta\mupn \p_\sigma \phi^\sigma-F\mupn\)=0\,.\nn
\ea
It is clear that the equation for $\phi^0$ involves a dangerous term with $\ddot \phi^0$ and that degree of freedoms seems a priori dynamical. However upon look, the kinetic matrix $\mathcal{A}_{ab}$ has a vanishing eigenvalue,
\ba
\mathcal{A}_{ab}=\mathcal{L}_m^{-3}\(
\begin{array}{cc}
-\(\phi^0{}'+\dot \phi^1\)^2 & \(\phi^0{}'+\dot \phi^1\)\p_\mu \phi^\mu \\
\(\phi^0{}'+\dot \phi^1\)\p_\mu \phi^\mu & -\(\p_\mu \phi^\mu\)^2
\end{array}\).
\ea
So even though all the equations of motion seem dynamical, they are not independent. This can be made manifest by constructing the following linear combination
\ba
\mathcal{C}&=&\p_\mu \phi^\mu \E_0-F_{0}^{\ \mu}\E_\mu\\
&=&\p_\alpha \phi^\alpha \p_{x} F_{01}-F_{01}\p_{x}(\p_\alpha \phi^\alpha )\nn
\ea
which involves no double time derivative term.
 There is therefore a constraint to be satisfied which prevents all the \stu fields to be independent propagating degrees of freedom. This constraint is precisely analogous to that obtained in the ADM language in 2 dimensions, \cite{deRham:2010kj} and then recently completely generally in four dimensions \cite{Hassan:2011hr}.

Saying it differently, if one were to solve the equation for $\ddot \phi^1$ using $\mathcal{E}_1$ and plug that solution back into $\mathcal{E}_0$, we would then not be able to solve for $\ddot \phi^0$ but would instead obtain the constraint mentioned above.

Or yet in other words, if we were to write $V^0=\dot \pi$, this would correspond to a perfectly well defined equation of motion for $\pi$.

\subsection{Including Gravity}

It is instructive to follow this 2d toy-model further and check the algebra of the whole constrained system.  For this we include gravity
 $\mathrm{d}s^2=-N^2 \mathrm{d} t^2+\gamma_{11}\(\mathrm{d}x^1 +N^1 \mathrm{d} t\)^2$.
The Lagrangian is then of the form
\ba
\mathcal{L}_m&=&2 N \sqrt{\gamma_{11}}\left[-1+\sqrt{(\mathcal{D}_-\phi^-)(\mathcal{D}_+\phi^+)}\right]\\
&=&2 N \sqrt{\gamma_{11}}\left[-1+\frac{1}{2\lambda} (\mathcal{D}_-\phi^-)(\mathcal{D}_+\phi^+) +\frac\lambda 2\right]\,,
\ea
where $\lambda$ is introduced as a Lagrange multiplier and we use light-cone coordinates,
\ba
\phi^\pm&=&\phi^0\pm\phi^1\\
\mathcal{D}_\pm&=&\frac{1}{\sqrt{\gamma_{11}}}\, \partial_x\pm\frac{1}{N}\left[\partial_t-N^1\partial_x \right]\,.
\ea
The associate conjugate momenta are then
\ba
P_\pm=\pm\frac{\sqrt{\gamma_{11}}}{\lambda}\, (\mathcal{D}_\mp \phi^{\mp})\,,
\ea
giving the following expression for the Hamiltonian
\ba
\mathcal{H}&=&-\frac{N}{\sqrt{\gamma_{11}}}\Big[-2 \gamma_{11}+P_+\partial_x \phi^+-P_-\partial_x \phi^-\Big]\\
&+&N^1\Big[P_+\partial_x \phi^++P_-\partial_x \phi^-\Big]\nn\\
&-&\hat\lambda \Big[\frac{P_+ P_-}{\sqrt{\gamma_{11}}} +\sqrt{\gamma_{11}}\Big]\,,\nn
\ea
where $\hat{\lambda}=\lambda N$.
From this expression it is clear that $N$, $N^1$ and $\hat{\lambda}$ are three Lagrange multipliers for three constraints, and that the Hamiltonian is pure constraint. Consistency of the constraint algebra will generate a fourth constraint, which is sufficient to remove all physical degrees of freedom. This is consistent with the expectation that a massive graviton in two dimensions has no degrees of freedom. The existence of additional constraints other than the two expected is an accident of two dimensions and does not follow in higher dimensions.

In the rest of the paper we focus on the four-dimensional scenario, where the counting goes as follows, the massless graviton carries two degrees of freedom, which added to the four \stu fields would lead to the usual five helicity modes of the graviton plus a sixth BD ghost. As we will see below, the theory of massive gravity presented here has a primary constraint which then generates a secondary constraint both of which ensure the absence of the BD ghost. In two dimensions there is are accidental tertiary and quaternary constraints which are necessary to remove all the degrees of freedom of the \stu fields.


\section{General Formalism}
The previous configuration shows explicitly why even though $\dot \phi^0$ (or $\ddot \pi$) enters quadratically in the action, its equation of motion is not independent of the other \stu fields, so not all of them are dynamical, and the BD ghost is hence absent. Now to go further we show how the same result remains true when all \stu fields are excited, and consider an arbitrary potential, \cite{deRham:2010kj}
\ba
\mathcal{L}_m=\mathcal{L}^{(2)}_{\rm der}(\K) + \alpha_{3} \mathcal{L}^{(3)}_{\rm der}(\K) +
\alpha_{4}  \mathcal{L}^{(4)}_{\rm der}(\K)
\ea
with $\K^\mu_\nu = \delta^\mu_\nu - \sqrt{\partial^\mu \phi^a \partial_\nu\phi^b\eta_{ab}}$ and
\beq
\label{L2der0}
\mathcal{L}^{(2)}_{\rm der} &=&[\K]^2-[\K^2]\,,\nn\\
\label{L3der}
\mathcal{L}^{(3)}_{\rm der}&=&[\K]^3-3 [\K][\K^2]+2[\K^3]\,,\\
\label{L4der}
\mathcal{L}^{(4)}_{\rm der}&=&[\K]^4-6[\K^2][\K]^2+8[\K^3]
[\K]+3[\K^2]^2-6[\K^4]\,.\nn
\eeq
 We then define the associated energy-momentum tensor as
 \ba
 X\mn&=&\K g\mn -\K\mn\\
 &+&(1+3\alpha_3)\(\K\mn^2-\K \K\mn+\frac 12 \([\K]^2-[\K^2]\)g\mn\)\nn\\
&+&(\alpha_3+4\alpha_4)\Bigg(\K\mn^3-\K \K\mn^2+\frac 12 \K\mn \([\K]^2-[\K^2]\)\nn\\
&-&\frac 16 ([\K]^3-3[\K][\K^2]+2[\K^3])g\mn\Bigg)\,,\nn
 \ea
 such that the equations of motion are easily expressed as
 \ba
\mathcal{E}_a = \p_\mu X^\mu_{\, a}=0\,.
 \ea
 The kinetic matrix is then
 \ba
 \mathcal{A}_{ab}=\frac{\partial \dot X^{0}_a}{\partial \ddot \phi^b}
 =\frac{\partial X^{0}_a}{\partial \dot \phi^b}.
 \ea
To prove the absence of ghost, it is sufficient to show that $\det \mathcal{A}=0$.

\subsection{Fourth-order expansion}
 For simplicity, we write in this section the \stu fields as $\phi^a=x^a+V^a$ and address the question order by order in powers of $V^a$. We will stop the analysis at quartic order in $V^a$ (in the Lagrangian, cubic order in the equations of motion) as it is at that order that potential problematic terms have been found previously. Once again we work in the locally inertial frame (as we show later, including gravity bears no effect on the result). Setting $\alpha_3=\alpha_4=0$, the equations of motion are then
 \ba
 \E_\nu=\p^\mu \(\mathcal{F}^{(1)}\mn+\mathcal{F}^{(2)}\mn+\mathcal{F}^{(3)}\mn+\cdots\)=0\,,
 \ea
 where $\mathcal{F}^{(n)}\mn$ is $n^{\rm th}$ order in $V^a$:
 \ba
 \mathcal{F}^{(1)}\mn &=&\mathcal{F}\mn\\
 \mathcal{F}^{(2)}\mn &=&\frac14 \Sigma\mn^2-\p_\mu V_\nu (\p_\alpha V^\alpha)+\frac 12 ((\p_\alpha V^\alpha)^2-(\p_\alpha V_\beta)^2)\nn\\
 \mathcal{F}^{(3)}\mn &=&-\frac18 \Sigma\mn^3-\frac 14 \Sigma\mn((\p_\alpha V_\beta)^2-\p_\alpha V_\beta \p^\beta V^\alpha)\nn \\
 &+&\frac 14 \Sigma_{\mu\alpha}\(\p_\beta V^\alpha\p^\beta V_\nu+\p^\alpha V_\nu \p_\beta V^\beta\)\nn\\
 &+&\frac 14 \mathcal{F}_{\mu\alpha}\p_\nu V^\alpha \p_\beta V^\beta+\eta\mn\Big[\frac{1}{12} (\p_\alpha V^\alpha)^3\nn\\
 &-&\frac 14 \p_\alpha V^\alpha \p_\beta V_\gamma \p^\gamma V^\beta+\frac 18 \p^\gamma V^\beta \p_\alpha V_\gamma \p^\alpha V_\beta\nn\\
 &-&\frac{1}{12}\p^\gamma V^\beta \p_\alpha V_\gamma\p_\beta V_\alpha\Big]\,,
 \ea
with $\Sigma\mn=\p_\mu V_\nu+\p_\nu V_\mu$. Here again, while neither $\mathcal{F}^{(1,2)}_{0\mu}$ depend on $\dot \phi^0$, the next order correction $\mathcal{F}^{(3)}_{0\mu}$ has an explicit dependence on $\dot \phi^0$. However one can explicitly compute the determinant of the kinetic matrix which turns out to vanish to that order in the expansion, $|\mathcal{A}|=\mathcal{O}((\p V)^3)$, so once again, only three of the four \stu scalars are dynamical. In this case, the associated constraint is expressed as
\ba
\mathcal{C}=\E_0+\(\frac 12 \mathcal{F}_{0i}-\frac 18 \Sigma_{0i}^2+\frac 12 \p_\alpha V_0 \p_i V^\alpha\)\E_i=0\,,
\ea
and involves no double time derivatives.

\subsection{Non-trivial background}

 It is clear from the previous expansion, that $V^0$ seem to acquire a kinetic term at quartic order in the expansion, or in other words, when expanding around a non-trivial background. To make this more explicit, and to connect with \cite{slavapaper2}, we give up the LIF in what follows and consider a specific example where the metric has a non-trivial background. Namely, we consider the following configuration, $\mathrm{d}s^2=-\mathrm{d} t^2+\delta_{ij}\(\mathrm{d}x^i +2 \ell^i \mathrm{d} t\)\(\mathrm{d}x^j +2 \ell^j \mathrm{d} t\)$. It is unclear whether such a background is stable and could be generated naturally in a physical system, as is explained in  \cite{deRham:2011pt}. In particular we know from the ADM formalism, that the shift cannot be excited without simultaneously exciting the 3d metric, \cite{deRham:2010kj,Hassan:2011hr}.
However, for the purpose of this work we follow here the same procedure as \cite{slavapaper2} and expand around this given background. We therefore focus on the quadratic fluctuations of the \stu field around this metric configuration, and to start with, we only keep track of the time derivatives (spatial derivatives are introduced later).

 \subsubsection{Primary Constraint}

 The potential is then of the form, (to all orders in $\ell^i$)
 \ba
 \label{Ubackground}
 \mathcal{L}_m&=&-\frac{1}{2\sqrt{1-\ell_k^2}}\Big[\, \eta\mn \dot V^\mu\dot V^\nu+\frac{( \dot V^0+\ell_i \dot V^i)^2}{1-\ell_k^2}\, \Big]\\
 &=&\frac 12 \dot V_i^2+ \ell_i\dot V^i\dot V^0+2 \ell_i\ell_j\dot V^i \dot V^j+\frac 14 \ell_i^2(\dot V^0)^2+\cdots \nn\,,
 \ea
and it is clear from the last term $\ell_i^2(\dot V^0)^2$ that $V^0$ seems to acquire a kinetic term. However the Hessian matrix associated with the kinetic terms of the \stu in  \eqref{Ubackground} is clearly degenerate,
\ba
\mathcal{A}_{ab}\propto\(\begin{array}{cc}
\ell_k^2 &  \ell_i\\
 \ell_j & \ell_i \ell_j+(1-\ell_k^2)\delta_{ij}
\end{array}\) \Longrightarrow \det \mathcal{A}=0\,,
\ea
which can be easily seen by setting $\dot W^i=\dot V^i+\ell^i\, \dot V^0$, such that the Lagrangian only depends on $\dot W^i$,
\ba
\mathcal{L}_m=-\frac{1}{2\sqrt{1-\ell_k^2}}\Big[(\dot W^i)^2+\frac{(\ell_i \dot W^i)^2}{1-\ell_k^2}\Big]\,,
\ea
and it is clear that only three of the four \stu fields are dynamical.
We emphasize however that this change of variable is not required to count the number of dynamical degrees of freedom or to see the constraint,
\ba
\mathcal{C}=P_0-\ell^i P_i=0\,,
\ea
where $P_\mu$ is the conjugate momentum associated with $V^\mu$, and we neglect spatial derivatives.

 \subsubsection{Secondary Constraint}

So far, we have neglected any space-like derivatives as the previous analysis was sufficient to prove the existence of a primary constraint, and hence already remove half a degree of freedom, already implying the absence of ghost.  However to study the algebra of the constraint system, and the existence of a secondary constraint, it is important to re-include space-like derivatives. When doing so, one can easily check that the canonical momenta associated with the fields $V^\mu$ are then
\ba
P_0&=&\ell^i \mathcal{F}_{i0}-\ell_k^2 \dot V^0+2 \ell^i \ell^j \p_i V_j-\ell_k^2 \p_i V^i\\
P_i&=& \mathcal{F}_{i0}-\ell_i \dot V^0-2 \ell_i \p_k V^k+\frac 12 \ell^k \p_i V_k\\
&& +\frac 52 \ell^k \p_k V_i+\ell_i \ell^k \p_k V^0-\frac 12 \ell_j^2 \p_i V^0\nn\\
&&-\frac 12 \ell_k^2 \p_0 V_i -\ell_i \ell^k \dot V_k\,, \nn
\ea
where for simplicity we stop the expansion at quadratic order in $\ell$.
It is then straightforward to check that the conjugate momenta are not independent but are related in the following way:
\ba
\label{constraint}
\mathcal{C}=P_0-\ell^i P_i-\ell_j^2\ \p_i V^i+\ell^i \ell^j \p_i V_j=\mathcal{O}(\ell^3)\,.
\ea
This relation, can still be regarded as a primary constraint as it
appears at the stage of determining the momenta.
Furthermore, requiring the conservation of the constraint \eqref{constraint}, leads to the secondary constraint $\mathcal{C}_2$,
\ba
\label{C_2}
\mathcal{C}_2&\equiv&\{\mathcal{C}, \mathcal{H}\}=0 \\
&\supset&\p_i P^i+\ell_i\p_jF^{ij}+\ell_k^2\p_i P^i+\cdots\,.\nn
\ea
To summarize, the Hamiltonian  is accompanied by the two constraints, \eqref{constraint} and \eqref{C_2}.
These two can be used to eliminate one of the coordinate-momentum
pair (for instance $(V^3,P_3)$, although any other pair is equivalently acceptable).
Hence, the system propagates three
physical degrees of freedom. In particular, from the primary
constraint \eqref{constraint} one can express $P_3$ in terms of $P_0,P_1,P_2,V^1,V^2,V^3$,
and from the secondary constraint \eqref{C_2} one can express $V^3$
in terms of $P_1, P_2, V^1, V^2, V^3$. Hence, the physical Hamiltonian
depends only on the unconstrained variables:
\ba
\mathcal{H}_{\rm phys}=\mathcal{H}_{\rm phys}(P_0,P_1,P_2,V^0,V^1,V^2)\,.
\ea

\section{Outlook}
The theory of massive gravity as proposed in \cite{deRham:2010kj} has been shown to be free of ghost both in the decoupling limit \cite{deRham:2010ik} and to fourth order in perturbations in the ADM formalism. Recently, Hassan and Rosen have advanced
the analysis further and shown the absence of ghosts to all orders in the ADM language, \cite{Hassan:2011hr}. While this result is in complete consistency with what was found in \cite{deRham:2010ik,deRham:2010kj} recent works seem to suggest the existence of a ghost either at quartic \cite{slavapaper2}, or even worse at cubic order, \cite{Folkerts:2011ev} in the \stu language, \footnote{To be precise, Ref.~\cite{Folkerts:2011ev} uses a similar but not identical methodology to diagnose the presence of a ghost by looking for higher derivative interactions of the helicity zero mode. Although the present article is concerned with \stu decomposition, it is clear that for similar reasons the analysis of \cite{Folkerts:2011ev} is incomplete, since within the helicity formalism, extra constraints will arise (as is inevitable for consistency with the ADM formalism), and these constraints will be seen to remove the apparent ghost.}.
If the conclusions of these works were correct one should seriously investigate why the ADM and \stu language lead to so different answers. In this work we have filled the missing gap between both languages and shown the existence of a constraint among the four \stu fields
which is responsible for the absence of ghosts. We perform this analysis to all orders in specific toy-models,  as well as
to quartic order in the full-fledged theory. In all these cases we show that the \stu analysis is in complete consistency
with the ADM language \cite{deRham:2010kj,Hassan:2011hr} and proves the absence of the BD ghost in this family of massive gravity
theories.

{\bf \emph{Acknowledgements:}} We would like to thank S.~Dubovsky and D.~Pirtskhalava for useful discussions.
CdR is funded by the SNF and the work of GG was supported by NSF grant PHY-0758032.




\begin{thebibliography}{99}


 \bibitem{Fierz:1939ix}
  M.~Fierz and  W.~Pauli,
  Proc.\ Roy.\ Soc.\ Lond.\  A {\bf 173}, 211 (1939).


\bibitem{vDVZ}
H.~van Dam and M.~J.~G.~Veltman,
  Nucl.\ Phys.\  B {\bf 22}, 397 (1970);
V.~I.~Zakharov,
  JETP Lett.\  {\bf 12} (1970) 312
  [Pisma Zh.\ Eksp.\ Teor.\ Fiz.\  {\bf 12} (1970) 447].



\bibitem{Arkady}
  A.~I.~Vainshtein,
  Phys.\ Lett.\  B {\bf 39}, 393 (1972);


 \bibitem{Boulware:1973my}
  D.~G.~Boulware and S.~Deser,
  Phys.\ Rev.\  D {\bf 6}, 3368 (1972).

\bibitem{ArkaniHamed:2002sp}
N.~Arkani-Hamed, H.~Georgi and M.~D.~Schwartz,
Annals Phys.\  {\bf 305}, 96 (2003).

 \bibitem{Creminelli:2005qk}
  P.~Creminelli, A.~Nicolis, M.~Papucci and E.~Trincherini,
  JHEP {\bf 0509}, 003 (2005).
  [arXiv:hep-th/0505147].

\bibitem{Gabadadze:2009ja}
  G.~Gabadadze,
  Phys.\ Lett.\  B {\bf 681}, 89 (2009)
  [arXiv:0908.1112 [hep-th]].


 \bibitem{deRham:2009rm}
  C.~de Rham,
  Phys.\ Lett.\  B {\bf 688}, 137 (2010)
  [arXiv:0910.5474 [hep-th]].


\bibitem{deRham:2010gu}
C.~de Rham and G.~Gabadadze,
  Phys.\ Lett.\  B {\bf 693}, 334 (2010)
  [arXiv:1006.4367 [hep-th]].

\bibitem{deRham:2010ik}
  C.~de Rham, G.~Gabadadze,
  Phys.\ Rev.\  {\bf D82}, 044020 (2010).
  [arXiv:1007.0443 [hep-th]].


\bibitem{deRham:2010kj}
  C.~de Rham, G.~Gabadadze, A.~J.~Tolley,
  Phys.\  Rev.\  Lett.\  106, {\bf 231101} (2011).
  [arXiv:1011.1232 [hep-th]].


\bibitem{Hassan:2011hr}
  S.~F.~Hassan, R.~A.~Rosen,
  [arXiv:1106.3344 [hep-th]].


\bibitem{Nieuwenhuizen}
T.~M.~Nieuwenhuizen,
  arXiv:1103.5912 [gr-qc].


\bibitem{Hassan:2011vm}
  S.~F.~Hassan, R.~A.~Rosen,
  [arXiv:1103.6055 [hep-th]].

\bibitem{Koyama}
K.~Koyama, G.~Niz and G.~Tasinato,
  arXiv:1104.2143 [hep-th].


\bibitem{slavapaper1}
  L.~Alberte, A.~H.~Chamseddine and V.~Mukhanov,
  JHEP {\bf 1104}, 004 (2011)
  [arXiv:1011.0183 [hep-th]].



\bibitem{slavapaper2}
  A.~H.~Chamseddine, V.~Mukhanov,
  [arXiv:1106.5868 [hep-th]].


\bibitem{deRham:2011pt}
  C.~de Rham, G.~Gabadadze, A.~J.~Tolley,
  [arXiv:1107.0710 [hep-th]].


\bibitem{Folkerts:2011ev}
  S.~Folkerts, A.~Pritzel and N.~Wintergerst,
  arXiv:1107.3157 [hep-th].



\end{thebibliography}
\end{document}